\documentclass[useAMS,usenatbib]{mn2e}
\usepackage{graphicx}
\usepackage{txfonts} 

\newcommand{\aap}{A\&A}
\newcommand{\aapr}{A\&A Rev.}
\newcommand{\mnras}{MNRAS}
\newcommand{\apj}{ApJ}
\newcommand{\aj}{AJ}

\newcommand{\araa}{ARA\&A}
\newcommand{\nat}{Nature}
\newcommand{\msait}{Mem. S.A.It.}
\newcommand{\pr}{Phys. Rep.}

\catcode`\@=11
\def\gsim{\ifmmode{\mathrel{\mathpalette\@versim>}}
    \else{$\mathrel{\mathpalette\@versim>}$}\fi}
\def\lsim{\ifmmode{\mathrel{\mathpalette\@versim<}}
    \else{$\mathrel{\mathpalette\@versim<}$}\fi}
\def\@versim#1#2{\lower 2.9truept \vbox{\baselineskip 0pt \lineskip
    0.5truept \ialign{$\m@th#1\hfil##\hfil$\crcr#2\crcr\sim\crcr}}}
\catcode`\@=12
\def\msun{\hbox{$M_\odot$}}

\def\yr-1{\hbox{${\rm yr}^{-1}$}}

\def\rsun{\hbox{$R_\odot$}}

\def\t9{\hbox{$t_9$}}

\def\m*{\hbox{$M_{\rm stars}$}}

\def\ho{\hbox{$H_\circ$}}
\def\h50{\hbox{$\ho /50$}}

\begin{document}

%\title{The Formation of Globular Clusters and their Multiple Populations: Where do we Stand?}
\title{The Formation of Globular Clusters as a Case of Overcooling}
\author[Alvio Renzini,  Anna F. Marino and Antonino P. Milone]{Alvio Renzini$^{1}$\thanks{E-mail: alvio.renzini@inaf.it}, Anna F. Marino$^{2}$\thanks{E-mail: anna.marino@inaf.it } and Antonino P. Milone$^{3}$\thanks{E-mail: antonino.milone@unipd.it}\\ 
$^{1}$INAF - Osservatorio
Astronomico di Padova, Vicolo dell'Osservatorio 5, I-35122 Padova,
Italy\\
$^{2}$INAF - Osservatorio Astrofisico di Arcetri, Largo Enrico Fermi 5, I-50125 Firenze, Italy\\
$^{3}$Dipartimento di Fisica e Astronomia Galileo Galilei, Universit\`a di Padova, Vicolo dell'Osservatorio 3, I-35122, Padova, Italy}

\date{Accepted 2022 April 2. Received 2022 March 11; in original form 2021 December 14}
\pagerange{\pageref{firstpage}--\pageref{lastpage}} \pubyear{2002}

\maketitle
                                                            
\label{firstpage}

\begin{abstract}
Driven by recent observational findings, we select massive interactive binaries as the most suitable among the existing candidates for producing the chemical  patterns typical of multiple populations of Galactic globular clusters. Still, to avoid supernova contamination we are further driven to endorse the notion that above a critical mass stars fail to produce supernova events, but rather eventually sink into black holes without ejecting much energy and heavy metals. This assumption has the attractive implication of suppressing star formation feedback for some 5--10  million years, in practice leading to runaway star formation,  analog to {\it overcooling} that in absence of feedback would have turned most baryons into stars in the early Universe. Under such conditions, multiple episodes of stars formation, incorporating binary star ejecta from previous episodes, appear to be unavoidable, thus accounting for the ubiquity of the multiple population phenomenon in globular clusters.
\end{abstract}


\maketitle

\section{Introduction}
\label{intro}
Chemical inhomogeneities among stars within Galactic globular clusters (GC) have been known to exist since a long time (e.g., \citealt{kraft79}) but it was only with the discovery of highly helium enriched sub-populations (\citealt{bedin04,piotto07}) that the complexity of the phenomenon began to unfold, hence triggering  a surge of photometric and spectroscopic studies, along with attempts at understanding how all this came about.  Hubble Space Telescope (HST) multiband, deep imaging \citep{piotto15} and Very Large Telescope (VLT) multiobject spectroscopy \citep{gratton12} then revealed an extremely rich phenomenology, more and more difficult to interpret, thus making the formation of GCs  with their multiple populations a major unsolved problem in astrophysics. The common pattern exhibited by GCs is evidence for CNO- and {\it p}-capture-processed material
dominating the surface composition of a major fraction of cluster stars, then called second generation (2G), whereas the remaining cluster stars, dubbed first generation (1G), show a surface composition similar to that of field stars of the same metallicity. This 1G/2G nomenclature implies that at least two distinct episodes of star formation took place, a point on which we shall return later. Moreover, what we see is a great deal of variance from cluster to cluster, in the number of stellar generations and in complexity of chemical patterns, having made more and more frustrating attempts at composing the puzzle.

In this paper, we let ourselves to be driven by recent observational progress in trying, in a bottom up approach, to sort out a GC formation scenario that should not violate the observational constraints. In Section  \ref{sec:data} we summarize major  advances gathered in the last few years, thanks to photometric and spectroscopic observations. In Section \ref{sec:where} we argue in favor of the idea according to which all metal-poor GCs formed inside dwarf galaxies, well before the main body of the Galaxy was built, an epoch that instead coincided with the formation of the stellar halo. In Section \ref{sec:bursts} we argue that current observations dictate that the multiple population phenomenon  is the result of multiple, successive episodes of star formation, leading to multiple stellar generations inside each GC progenitor. In Section \ref{sec:who} we discuss pros and cons of the various candidate polluters of CNO/{\it p}-capture processed material needed to form the 2G stars, as so far proposed. Among them, we single out massive interactive binaries as the most promising ones, an option that requires the additional assumption  that massive stars above a critical mass do not explode as supernovae, failing to eject much energy and metals. In Section \ref{sec:what} we try to put together the pieces of the puzzle, as a sequence of plausible events, that culminate with the emphasis on the inevitability of multiple stellar generations in absence of supernova feedback. In Section \ref{sec:rich} we acknowledge that metal rich GCs cannot have formed inside dwarf galaxies, but rather did so inside the forming Galactic bulge, where conditions leading to GC formation were however much rarer there than in dwarf galaxies. Finally, in Section \ref{sec:summary} we wrap up our scenario of globular cluster formation and conclude.

\section{Main observational progress in the last five years}
\label{sec:data}
The "HST UV Legacy Survey of Galactic Globular Clusters" \citep{piotto15} has provided an exquisite, homogeneous documentation of the multiple population phenomenon for 57  GCs spanning a wide, representative range of masses, metallicities and 1G/2G phenomenology. It did so by combining four filters from the UV to the $I$ band, namely, F275W, F336W, F438W and F814W. The three bluest passbands include the OH, NH and CN+CH molecular bands, respectively, thus allowing to distinguish among stars with different degrees of CNO processing having taken place.
A color-pseudo-color combination of these passbands has resulted in plots, dubbed Chromosome Maps (ChM), that most effectively allow us to identify 1G and 2G stars in each cluster along with
sub-populations within both the 1G and 2G ones \citep{milone17}. From this study, three main results are worth recalling here: 1) the fraction of 2G stars, i.e., 2G/(1G+2G), increases with cluster mass from $\sim\!\! 50$ per cent for $\lsim 10^5\msun$ clusters to over 80 per cent in $>10^6\msun$ clusters, i.e., that of 2G is the main GC formation phase; 2) in 10 out of the 57 GCs the ChM is split, with two nearly parallel sequences and therefore GCs where classified in two groups: Type I clusters showing single 1G and 2G sequences and Type II clusters with separate blue and red sequences; and 3)
the 1G locus on the ChM of many clusters is appreciably spread in F275W-F814W color, signaling that even the 1Gs are not fully chemically homogeneous.

Further progress in the characterization of the multiple population phenomenon included the following. Estimates of the maximum enhancement of the  helium content $\delta Y_{\rm max}$ of 2G stars with respect to 1G stars, found to range from just $\sim\!\!0.01$ for $\lsim\! 10^5\msun $ GCs up to over $\sim \!\!0.08$ for $>10^6\msun$ GCs \citep{milone18}. In this study the effects on the ChMs was explored when the  abundance of the elements C, N, O, Mg and He are changed, one at a time.  It turned out that the helium vector runs perfectly parallel to the 1G sequence suggesting that even 1G stars in some GCs could exhibit a spread in helium abundance, an idea expanded  by \cite{lardo18}. Indeed, at first sight 1G sequences appear dominated by helium variations without a concomitant
CNO processing (as if due solely to $pp$-chain processing), whereas the 2G sequences result from a combination of both helium and nitrogen increases (as expected from CNO-cicle processing).
However, Milone et al. (2018), having explored various possible scenarios, concluded that there appears to be no plausible stellar process that can produce sizable amount of helium without accompanying it with a large increase in nitrogen, and therefore the 1G color spread had to be due to something else than helium. The 1G ChM sequence  of the cluster NGC 3201 was then spectroscopically explored by \cite{marino19a} finding a $\sim\!\!0.1$ dex spread in [Fe/H] with a few extreme stars showing radial velocity variations, thus indicating that the 1G spread is due to a combination of modest iron variations with binary stars contributing to the color spread.  Binarity was further investigated in five GCs by \cite{milone20}, quantifying the frequency of 1G+1G, 2G+2G as well as 1G+2G main sequence binaries.

Another important piece of evidence was added by extending the search for multiple populations below the main sequence ``knee" produced by the formation of the H$_2$ molecule, and well down into the fully convective M dwarf regime. This needed to add near-IR passbands, namely F110W and F160W (\citealt{milone19} and references therein). Thus, in the GC NGC 6752, below the knee the main sequence splits in at least three diverging sequences which are due to differences in water vapor blanketing in the F160W band. Thus, 2G stars, which are oxygen depleted, suffer less H$_2$O blanketing than 1G stars which have ``normal" oxygen abundance and therefore are bluer in the F110W-F160W color compared to 1G stars. The same approach has been recently extended to ten GCs, confirming the split of the lower main sequence and demonstrating that (in the clusters with deeper data) the slope of the mass function is the same for all 1G and 2G populations and their proportions are  consistent with those exhibited by 1G and 2G upper main sequence and red giant stars \citep{dandoglio22}.

Finally, extensive chemical tagging of the ChMs of 29 GCs was accomplished by \cite{marino19b} using virtually all stars in the HST ChMs having chemical abundance analysis from own previous and literature studies. Besides confirming a $\sim \!\!0.1$ dex spread in [Fe/H]   among 1G stars, this study has shown that the split ChM sequences in Type II clusters is due to a difference in [Fe/H] that can be as high as $\sim \!\!0.3$ dex in extreme cases. Interestingly, Marino et al. (2919b) have constructed a ``universal" ChM by stacking the ChM of individual clusters after rescaling them to virtually eliminate the effects of different metallicities. This universal ChM shows narrow, well separated  1G and 2G sequences, with different clusters occupying different portions of these sequences.
This demonstrates that the underlying physical processes, of chemical processing and star formation, ought to have an intrinsic underlying order,  as opposed to once favored mere stochasticity
\citep{bastian15}. Still, it remains to be understood how this order was established in Nature.

%section

\section{Where Did Globular Clusters Form?}
\label{sec:where}
The GC with the most complex display of multiple populations is undoubtedly $\omega$ Centauri, which hosts at least 15 distinct stellar populations \citep{bellini17}.
Though at that time such complexity was not known yet, \cite{bekki03} proposed that $\omega$ Cen was the remnant nuclear cluster of a now dissolved dwarf galaxy. We now wish to endorse an Ansatz such that not just $\omega$ Cen, but virtually all GCs formed inside dwarf galaxies (as first proposed by \citealt{searle78}), perhaps with the exception of the more metal rich ones. Indeed,  the association of GCs with dwarf galaxies (as in the case of M54 and Sagittarius) and with Gaia streams is now being widely pursued (e.g., \citealt{massari19}). On the other hand, even the most metal poor GCs had to form in an environment in which the metallicity had reached about 1/100 solar, as expected for dwarf galaxies at over 12.5 Gyr lookback time, or $z\gsim 3$. In this picture GCs did not hurry to separate themselves from the womb where they formed. This may have taken several Gyr to do so. While waiting for the Galaxy to grow, young GCs  were still immersed in the dense gas of the host dwarf that worked as a tamper holding the stellar ejecta within the new born cluster and possibly feeding it with stellar ejecta from stars in the dwarf itself. 

In any event, GCs are not pristine, though  at least the metal poor largely predate the Galaxy.
Often GCs are thought as having been accreted by the Galaxy, as if the Galaxy was already in place and the GCs arrived later, whereas it may well be that the actual late-comer is the bulk of the Galaxy itself, that was slowly built up by inflowing gas streams as currently understood \citep{dekel09}. Thus, as the Galaxy grew up, most dwarfs, but not all, have been tidally destroyed, then releasing their GCs.

GCs are extremely compact objects, with density of the order of $\sim \!\!10^7$ particles cm$^{-1}$ (now all in stars), i.e., many thousand times that of typical molecular clouds in today's Milky Way. 
Runaway gas cooling and gravitational collapse must have occurred to form them, but  it is less clear how gas was squeezed to such high densities {\it before} forming stars. On the other hand, besides being very dense, GCs are also extremely slow rotators \citep{sollima19}. It is possible that GCs can only form in deep local minima of the potential well created by the collapse of giant molecular clouds and in minima of the interstellar medium  (ISM) vorticity, which may explain why (thin) disk GCs do not exist. Otherwise, a mechanism should be invented as to remove angular momentum with extremely high efficiency from giant molecular clouds. 

Tantalizing direct evidence for a GC forming inside a dwarf at $z\sim \!6$ was recently obtained from a highly lensed  dwarf galaxy with a size of $\sim \!\!400$ pc and a stellar mass of $\sim\!\!2 \times 10^7\msun$ which  hosts a compact, unresolved nucleus with $R_{\rm e}<13$ pc and a stellar mass of $\sim \!10^6\msun$ \citep{vanzella19}.
The extent to which this phenomenon may be ubiquitous at high redshift could soon be explored with JWST (e.g., \citealt{pozzetti19}).

In any event, if forming GCs and their parent dwarfs were tightly interconnected, then a fraction of the dwarf itself may have contributed CNO/{\it p}-capture processed materials for the production of multiple stellar generations in the clusters. This may help solving the infamous ``mass budget problem" affecting all scenarios for the origin of multiple populations, i.e., the fact that, in each cluster, the present 1G stars fall short by about a factor of $\sim\!\!10$ from  having produced all the stuff needed to make the 2G ones (e.g., \citealt{renzini15}). Thus, for this to happen the part of a dwarf producing the 1G stars feeding the nascent  GC should have a mass roughly 10 times that of a present-day GC, i.e., a few per cent of the dwarf mass. Indeed, GC masses are typically in the range $10^5-10^6\,\msun$, whereas dwarf masses reach up to several $10^8\,\msun$. This dwarf galaxy picture of a progenitor GC is radically different from one in which progenitors are seen as just more massive, compact clusters, i.e., as a naked, massive GC. Objects of this later kind are hard to destroy, whereas most dwarfs hosting GCs can survive only as long as the main body of the Galaxy has not been assembled yet, and therefore tidal forces are still relatively weak. Thus, as the dwarf stars are tidally stripped, so are most of the  stars of the 1G generation that provided the material to form the 2G stars.

One additional option offered by the dwarf precursor scenario is that two, or more clusters may have formed within the dwarf, at slightly different times, hence with different metallicities, and that later in-spiralled  to merge. For example, such cluster merging was invoked for the Type II cluster NGC 1851 \citep{bragaglia21}.

%section

\vspace{-0.3 truecm}
\section{Why multiple episodes of star formation?}
\label{sec:bursts}
At early times, in the 'seventies and 'eighties, there was some reluctance to abandon a paradigm that was looking at GCs as prototypical ``simple stellar populations",  or SSPs. The  origin  of the chemical inhomogeneities was hopefully sought in internal processes within individual stars, or  appealing to accretion from a contaminated interstellar medium (e.g., \citealt{kraft79, dantona83, renzini83, iben84}). Yet, several arguments support the notion that it is inescapable to have distinct, successive episodes of star formation within forming GCs. First came the discovery that ``chemical anomalies" (as they were called) are not confined to red giants but extend down to the main sequence \citep{hesser80}. This excluded the possibility that the chemical anomalies could be established in red giants via some sort of unconventional mixing. As we argue next, modern data further emphasize the case for multiple generations inside individual GCs. 

Accretion onto pre-existing stars cannot work for several reasons. There is indeed a clear 1G-2G separation in most GCs and it is hard to understand why some 1G stars would have escaped completely from accreting material while others would have accreted a lot and be turned into 2G ones. Moreover, discreteness is not confined to 1G vs. 2G, but the ChM of many GCs shows that the 2G sequence is made of several disjoint clumps, again hard to understand how such clumps would have been generated by an accretion process. Moreover, in several GCs the most extreme 2G population is made of stars which are extremely oxygen depleted, by up a factor of $\sim \!\!10$ with respect to 1G stars in the same cluster. These stars are also strongly helium enhanced, hence have higher molecular weight compared to 1G stars. Accretion onto pre-existing 1G stars of oxygen-depleted/helium-enriched material cannot account for these stars, because the Rayleigh-Taylor instability would quickly  lead to mixing of the accreted, higher molecular weight material with the rest of the star, then washing away the oxygen depletion. Hence, highly oxygen-depleted stars ought to have formed out of almost pure 1G stellar ejecta. Finally, the identity of the mass functions of 1G and 2G lower main sequences \citep{dandoglio22} is hard to accommodate in an accretion scenario, where accretion is expected to favor more  massive stars.
All these issues are instead quite naturally accounted if GCs formed through a series of star formation  events (bursts), from two to several. Problems exists but are elsewhere for this scenario, whereas the accretion hypothesis has never been confronted with the actual complexity of the data.

%section

\vspace{-0.752 truecm}
\section{What Stars Have Produced the 2G Raw Material?}
\label{sec:who}
Type I GCs are quite homogeneous as far as the iron abundance is concerned, with the exception of the mentioned $\sim\! 0.1$ dex spread in [Fe/H] among the 1G stars. 
The presence of a $\sim \!\!0.3$ dex difference in [Fe/H] between the parallel blue and red ChM sequences of Type II GCs  testifies that some supernova contamination took place after the formation of the blue 2G sequence of these clusters and before/during the formation of their iron-enriched, red sequences. Yet, in both cases the amount of extra iron involved, in bulk solar masses, is relatively tiny: at most just a few per cent of the total iron released by core collapse supernovae (CCSNe) from the first generation \citep{renzini13, marino19b}. This means that both  kinds of 2G stars had to form either before or well after the explosion of the bulk of CCSNe produced by the first generation. This is known as the supernova avoidance problem \citep{renzini15}.  
Alternatively, supernova ejecta may have preferentially escaped from the forming GC, see later in  this section, but we consider less attractive this option as it would possibly exacerbate the mass 
budget problem if entraining other gas.

Supernova avoidance can be realized in two distinct ways, depending on the nature of the polluters, i.e., the  kind of stars that produced the CNO/{\it p}-capture processed material. If those were massive or supermassive stars, then the whole GC formation, with all 1G and 2G stars,  had to be a very fast process, lasting just a few to several million years, so that only a few massive stars had the time to turn into CCSNe. So, this timescale is dictated by the supernova avoidance constraint. 

Alternatively, if the polluters were intermediate mass AGB stars, one had to wait of order of a few $10^8$ yr, so to accumulate enough processed material, before starting to form the 2G stars.
Almost all the metals produced by the 1G CCSNe had to disappear completely for Type I clusters, while some 2 per cent being retained by Type II clusters as needed to form their ChM red sequence.

In all evidence, both scenarios are highly contrived. In the case of massive star polluters, binaries should play the main role \citep{demink09} as some 70 per cent of OB-Type stars are interacting binaries \citep{sana12}. Close binaries offer the additional advantage of quite naturally promoting fast rotation and deep envelope mixing during the core hydrogen burning phase of their components.
Then, prompt common-envelope phases of the binaries can release their whole envelopes, fully processed by CNO-cycle and {\it p}-capture processes, well before experiencing their core collapse. Moreover, the mass lost during common envelope phases is shed at low velocity, hence with minimal energy injection into the ISM, an aspect that may turn relevant, as we shall see later. For these reasons, in our opinion  close binaries offer a more attractive option over the single fast rotating massive stars (FRMS) alternative  \citep{krause13}, which can violate the GC-specific constraint$^3$\footnotetext[3]{GC specificity indicates that the multiple population phenomenon is typical of GCs but rare in the field.} \citep{renzini15} while an extruding disk does not appear to be an obvious place for star formation to occur. 

Supermassive stars (SMS, $\sim\!\! 10^4-10^5\msun$, \citealt{denissenkov14}) very well satisfy the GC-specificity constraint but encounter another serious difficulty. The mass of these SMS models cannot exceed this limit, otherwise their interior would be too hot for producing {\it p}-process elements in the observed mix. Thus, a few times $10^4\msun$ SMS cannot physically produce the $\sim\!\! 10^5\msun$ of excess helium now stored in the 2G stars of $\omega$ Cen \citep{renzini15}. To overwhelm this difficulty, \cite{gieles18} have proposed a scenario in which the SMS, once formed, keeps accreting stars at a rate similar to its wind mass loss, so there would be virtually no limit to the amount of material that the SMS could process. However, for this mechanism to work several conditions should be verified, including: a) central temperature and density must be high enough to timely produce {\it p}-processed material in the desired amount; b) the SMS must be fully convective$^4$\footnotetext[4]{Hence the SMS structure would be a polytrope of index $n=3$.}, in such a way that this material is continuously brought to the surface and shed to the ISM; and c) 
the radius of SMS should be very large, so that  stars are  captured at the required rate. These conditions can  be mutually exclusive. For example, in their scenario the authors assume the radius of a $10^5\msun$ SMS to be 30,000 $\rsun$, which implies a mean density of only $\sim 6\times 10^{-9}$ g cm$^{-3}$ or a central density of $\sim 3\times 10^{-7}$ g cm$^{-3}$  for a polytrope of index 3 \citep{cox68}.  For comparison, the central density of the Sun is $\sim 9$ orders of magnitude higher. It remains to be seen whether an SMS with these properties (mass, size, convection and {\it p}-process production)
could really exist, but we consider it unlikely. For certain, SMS stellar models satisfying all three constraints do not currently exist and, as it stands, the proposed SMS scenario unfortunately does not appear viable.

AGB stars were soon and for a long time regarded as the natural polluters for the production of multiple populations. Candidates would be stars in the mass range $\sim\!\! 3-8\, \msun$, hence with lifetimes of $\sim\! 30$ to 300 Myr. Some apparent mismatch concerning the detailed abundances, such as a predicted Na-O correlation in front of the observed anticorrelation, could be alleviated by either appealing to modification  within current uncertainties of some nuclear reaction rate \citep{renzini15}, or by tuning AGB models and their mass loss in order to have most of the fresh sodium to be ejected before being destroyed. So, chemistry may not be fatal for the AGB scenario. More difficult is to understand how to avoid that 1G supernova products be incorporated into 2G stars, given that at most $\sim\!\! 2$ per cent of them are so in Type II clusters and almost none in Type I clusters. If GCs form inside dwarfs, this would require to eject from the dwarf itself most 
($\sim\!\! 98$ per cent) of the CCSNe products while still retaining the AGB products and at least part of the initial ISM of the dwarf, a very contrived scenario. In principle, given the shallow potential well of the dwarfs, the supernova shock heated {\it bubbles} could manage to carve windows in the ISM and outgas from them with minor ISM entrainment. Existing simulation offer a contradictory panorama, with the supernova ejecta being either fully retained by the protocluster \citep{krause13}, or selectively fully ejected  \citep{tenorio16} or fully ejected along with the original ISM of the precursor \citep{dercole10}. All supernova ejecta should instead to be replaced by the materials lost by the AGB stars plus some original material having somehow escaped the supernova contamination. It appears quite unlikely that virtually all forming GCs would have followed this very specific sequence of events. This makes the weakness of the AGB option. 

However, also the massive binary star alternative has to face the supernova avoidance problem, an aspect that was not addressed in \cite{demink09}. Massive binaries filling their Roche lobes and shedding CNO/{\it p}-capture processed materials are indeed both co-spatial and time concomitant with massive stars, single and double alike, finally undergoing core collapse. Hard to imagine how Nature could separate the precious CNO/{\it p}-capture processed matter from the unwanted metals ejected by supernova explosions, given that both occupy the same space at the same time. The only escape, it seems to us, is to assume that above a certain stellar mass ($M_{\rm noSN}$) core collapse is not accompanied by supernova explosion and metal ejection, but the core ``silently" sinks into a black hole.  
This was postulated by \cite{krause13} precisely to ensure supernova avoidance in their FRMS scenario. Though independent arguments exist supporting this possibility (e.g., \citealt{sukhbold16,adams17,sander19,eldridge22} and references therein), it remains somewhat perplexing that a very massive star of several ten solar masses could {\it quietly} sink into a black hole without much energy release and ejection of processed materials.

Yet, as of today, the massive binary star option, with the {\it ad hoc} assumption that stars above a certain mass limits fail to produce a supernova event, appears to be more promising than other scenarios. It is based on the presumption that in an environment with extremely high gas densities ($\gsim 10^7$ cm$^{-3}$) multiple bursts of star formation are inevitable, the problem actually being how to stop it rather than how to go on and on with an extremely high efficiency of gas conversion into stars. Given the ubiquity of multiple generations among GCs, multiple bursts ought to be an inevitable outcome, rather than the result of a fortuitous  conspiracy.   This is further explored next.

%section

\vspace{-0.3 truecm}
\section{What was the sequence of events?}
\label{sec:what}
Here we explore further the scenario of GC formation based on three assumptions: 1) GCs (or at least the metal poor ones) formed inside dwarf galaxies; 2) 1G massive binaries produce the CNO/{\it p}-capture materials that are ejected during their common envelope events and then incorporated into 2G stars; and 3) massive stars above a certain limit (e.g., 20-40 $\msun$ [?]), no matter whether single or binary) do not end up with a metal producing supernova,  but sink directly into a black hole.

The first assumption has already been justified in Section 2. Thus, in this view progenitors of GCs are not just bigger GCs, but the host dwarfs themselves that can be much more massive than the hosted GC(s).  Therefore, binary star polluters to make 2G stars are not only those in the 1G inside the just formed GC, but also those in a suitable portion of the dwarf itself. This would solve the mass budget problem. The very high gas density inside the the forming GC and its surrounding regions within the dwarf galaxy should ensure rapid energy dissipation, such as that from the fast winds of massive stars, favoring further collapse of binary ejecta down into the GC. This process may not be continuous, but rather going through ups and downs in a series of bursts thus making the discrete multiple populations exhibited by many ChMs. In some case, the very first 2G stars may form from almost pure 1G ejecta, thus making the highly oxygen depleted 2G stars that are present in some cluster.

The duration of star formation depends on the assumed stellar mass above which no metal producing supernovae take place. This would be $\sim \!\!10$ Myr for $M_{\rm noSN}=20\,\msun$ or
$\sim\!\!5$ Myr for $M_{\rm noSN}=40\,\msun$. Notice that $M_{\rm noSN}$ cannot be too small, otherwise the global metal yield would become insufficient to account for the metal enrichment of galaxies \citep{krause13}. Nevertheless, 5 to 10 Myr appear to be a sufficient time to accommodate multiple stellar generations from an ISM whose chemical composition is evolving rapidly in response to the input from 1G stars of the young GC and the host dwarf. Most importantly, the absence of supernova explosions during this time, i.e., of feedback, can only help turning more gas into 2G stars. Indeed, in absence of feedback an {\it overcooling} radial flow appears to be inevitable, with virtually all gas being turned into stars, as long ago recognized it would have been so in the early Universe \citep{white78}. Therefore, the postulate of having no supernovae above 
$M_{\rm noSN}$ so to avoid metal pollution has also the additional, beneficial effect of suppressing most of feedback for $\sim\!\! 5-10$ Myr, hence facilitating the growth of young GCs with more and more 2G populations. Of course, suppressing supernova feedback does not eliminate feedback completely, because the effect of fast winds from hot, massive stars remains. This lesser feedback may introduce modulations in star formation rate during the 2G built up, then resulting in secondary bursts leading to the  multiple 2G populations  present in several GCs. Notice that in such an overcooling phase most, if not all, binary ejecta are converted/incorporated into 2G stars, thus leading to a major step towards solving the mass budget problem.

However, as no supernovae explode during this time there remains to explain why a $\sim\!\!  0.1$ dex spread in [Fe/H] exists among 1G stars \citep{marino19a}. The only option seems to be that the ISM out of which 1G stars formed was not perfectly mixed, which is possible but hard to prove independently.  Different is the case of the iron enrichment in Type II clusters \citep{marino19b}.
Indeed, it is conceivable that after the no supernova phase came to an end and the ISM started to be enriched in metals, then another burst of star formation could have taken place, then leading to the formation of both the 1G and 2G stars of the metal rich (red) sequences in the ChM of these clusters. As mentioned earlier, it is also conceivable that in some cases more than one GC formed within a given dwarf, with slightly different [Fe/H], and that they later merged to form a single GC \citep{bragaglia21}. Indeed, the extreme complexity of the ChM of $\omega$ Cen  \citep{milone17,marino19b} argues for multiple, separate star forming protoclusters that eventually merged together after forming their 1G and 2G stars and even while being enriched in iron.

Though still very contrived, this scenario seems able to qualitatively account for all the major observed properties of GC multiple populations. For example, the increase of the 2G fraction with cluster mass \citep{milone17} can be the natural result of the deeper potential well in which they resided, favoring the accumulation of more 1G ejecta from the dwarf 1G population and promoting a higher number of multiple bursts. 

Finally, most dwarfs where stripped away by tidal interactions as the main body of the Galaxy grew in mass. In this respect, one last constraint is set by the fact that in the field stars with 2G chemical patterns are very rare \citep{martell11}, dictating that most of the 2G stars formed inside the GC itself and were retained by it. This means that the various bursts of star formation had to take place well inside the already formed, young GC. Support to this possibility comes from the fact that 2G stars appear to be more centrally concentrated compared to 1G stars (e.g., \citealt{milone12}).

This overall scenario for GC formation is quite similar to the one advocated by \cite{elmegreen17}, with the main difference being the extension of the  time interval  without supernova explosions, from $\sim\!\! 3$ up to $\sim\!\! 10$ Myr. Suppressing supernova feedback for several million years has the effect of boosting the overall star formation efficiency (fraction of the gas turned into GC stars), allowing more 1G-processed material to become available for the formation of 2G stars. With only 3 Myr duration for the star formation phase, only binary components more massive
than $\sim\! 100\,\msun$ would have had time to evolve off the main sequence \citep{yusof22},
fill their Roche lobe, and deliver their processed envelopes. This would probably fall short 
of accounting for the required mass budget, whereas extending to a 10 Myr interval would allow 
stars down to $\sim\! 20\, \msun$ to contribute. A longer duration of this phase has also the effect 
of allowing a $\sim\! 30$ times wider volume inside the dwarf to contribute to the inflowing 
stream feeding the forming GC, again busting further the mass budget. This assumes that the starburst is not confined to the nascent GC, but at the same time extends over a suitable region around it. The $\sim\! 100$ pc wide 30 Doradus starbursting region 
in the LMC, with its central R136 cluster, is highly reminiscent of the proposed process of GC formation. Indeed, it has been seen as a nearby analog to GC formation at high redshift, with the 
formation of the central cluster  taking  place within an extended region that has been actively star forming over the last $\sim\!\! 10$ Myr  (\citealt{schneider18} and references therein). In summary, in this proposed scenario, GCs form over a $\sim\!\! 10$ Myr time interval, as the central peak of a starbursting region encompassing a mass some 10 times larger than the mass of the final GC, corresponding to a few percent 
of the total mass of the hosting dwarf galaxy.

%section

\vspace{-0.3 truecm}
\section{Do Metal Rich Globular Clusters Fit in this scenario?}
\label{sec:rich}

Metal rich globular clusters do not quite fit in this scenario. The most metal rich ones (NGC 6528 and NGC 6553) are $\sim\!\! 10-12$ Gyr old, almost solar in iron and slightly enhanced in 
[$\alpha$/Fe] \citep{ortolani95,montecinos21}. Given the mass-metallicity relation of high redshift galaxies (e.g., \citealt{erb06,kashino17}), clearly they did not form inside a dwarf. Rather, they must have formed while the Galaxy had already grown to some $10^{10}\,\msun$. Yet, clusters such as NGC 6388 and NGC 6441 ([Fe/H]=--0.5) exhibit  prominent 2G sequences in their ChM \citep{milone17, tailo17, bellini13}, whereas this is less obvious for NGC 6553 \citep{montecinos21}, which has near solar metallicity. Therefore, sites other than high redshift dwarf galaxies have been able to form GCs with their multiple generations. Giant star-forming clumps in $z\sim 2$ galaxies have been proposed as such possible sites \citep{shapiro10}, with clumps possibly migrating inward to built the bulge, where indeed metal rich GCs reside. Still, for this to happen along with multiple stellar generations, conditions similar to those prevailing in high-redshift dwarfs should have been present also in early disks/bulges  of more massive galaxies. This includes very high gas fraction and relatively low specific angular momentum of gas rich clumps, so to make possible a collapse to very high gas and stellar mass densities. 

In this respect, it is worth noting that the vast majority  of the $\sim\!\!150$ Galactic GCs reside in the galactic halo, which has a mass of $\sim 10^9\msun$, whereas only $\sim 15$ of them belong to the bulge \citep{barbuy98}, which has a mass of $\sim\!\! 2\times 10^{10}\msun$ \citep{valenti16}. Thus, the metal rich bulge was some 100 times less efficient in forming GCs than the metal poor stellar halo,
a known effect also common among giant elliptical galaxies \citep{harris02,lamers17}. Thus, it appears that conditions leading to GC formation were present also during the built up of the Galactic bulge, but such conditions were very much rarer in the bulge than in pre-galactic dwarfs.

On the other hand, multiple generations are not exclusive to ancient GCs formed at high redshift. Evidence for chemical patterns similar to those exhibited by Galactic GCs have indeed been reported for Magellanic Cloud GCs as young as $\sim 2$ Gyr, whereas so far no sign for them have been found in younger clusters \citep{martocchia18}. However, note that the filter combinations needed to construct the ChMs are generally unavailable for such clusters.

%section
\vspace{-0.5 truecm}
\section{Summary and Conclusions}
\label{sec:summary}

The ubiquity of the multiple population phenomenon in GCs indicates that the processes leading to them  must be virtually unavoidable during the formation of massive GCs, as opposed to a combination of fortuitous conditions. In search for such processes, we have proceeded bottom up, starting from the observational evidences, especially those accumulated in recent years.
We started endorsing the notion that GCs, at least the metal poor ones, formed at a very early stage inside dwarf galaxies pre-existing the main body of the Galaxy itself. We then argued that
the multiple population phenomenon is more readily understandable as a sequence of star formation events, leading to multiple stellar generations, as opposed to alternatives such as accretion onto pre-existing stars, as this latter scenario appears to be incapable of complying with major observational constraints.

We then briefly considered, one by one, the various stellar candidates for the production of the chemical material needed for the production of multiple populations. This included AGB stars, fast rotating massive stars, supermassive stars and massive interactive binaries. These latter ones, the binaries, have been singled out as those offering the most promising option.
However, an additional assumption became necessary to avoid major supernova contamination of the CNO/{\it p}-capture processed material gently shed by the binaries during their common-envelope events. We then came to adopt also the notion that, above a critical mass,  stars would fail to explode as supernovae ejecting lots of energy and metals, but rather sink into black holes without much display.

The lack of supernova explosions for a period of some 5 to 10 million years has the important consequence of suppressing most of the star formation feedback shortly after the formation of the first generation of GC stars. Under such conditions, nothing prevents the residual gas, together with the binary star ejecta, to keep flowing into the nascent GC while forming stars in a series of bursts, until supernovae  finally begin. 
This physical situation is analog to that leading to overcooling, that in absence of feedback would have turned all the baryons into stars in the early Universe. The attractive aspect of this scenario is that multiple stellar generation become unavoidable once the first GC generation has formed, thus accounting for the ubiquity of the multiple stellar population phenomenon. If this is the way GCs formed, then such a {\it delayed feedback} is likely to play a role in star formation in general. In particular, it would boost the star formation efficiency in dense regions, perhaps helping the formation of nuclear clusters and more, all possibilities whose exploration is beyond the scope of the present paper.

Yet, not all GCs may have formed inside a dwarf. Certainly not those in the Galactic bulge which are metal rich. We take this as evidence that conditions conducive to GC formation existed also during the major epoch of star formation in the bulge, some 10 Gyr ago \citep{renzini18}, such as extremely high gas density coupled with low vorticity. However, given the specific frequency of bulge and halo GCs (i.e., their number per unit halo/bulge mass) it appears that such conditions were $\sim\!\! 100$ times more rare in the bulge than they were in dwarfs that once dissolved left free their GCs and made the stellar halo of the Galaxy.

In this proposed scenario, a few observational facts still remain unaccounted. This is the case for the presence of lithium and/or some heavy s-process elements (such as barium) in a few second generation stars, especially among the red-sequence stars of Type II clusters \citep{marino19b}.  Lithium and s-process elements have been traditionally taken as the smoking gun for the intervention of AGB stars (e.g., \citealt{dantona19} and references therein). As red-sequence stars in Type II GCs experienced iron enrichment from supernovae, over $\sim 10$ Myr after the formation of the first generation, they may as well have included ejecta from massive AGB stars that started to enrich the ISM another $\sim\!\! 20$ Myr after the beginning of supernova explosions. 
Admittedly, this tentative solution is a bit contrived, but so complex is the overall multiple population phenomenon that some apparently minor detail may need a specific fix.

In conclusion, after having defended the AGB options for many years, given the observational evidences we and others have recently gathered, we came to consider massive interactive binaries as the most promising factories for the production of the CNO/{\it p}-capture processed material used to make second generation stars. We further advocate  the {\it ad hoc} assumption that massive stars (either single or binary) above a critical mass fail to explode as supernovae, hence for some 5 to 10 Myr failing to provide feedback to regulate and stop further star formation. We argue that during this period, in absence of supernova feedback, {\it overcooling} leading to repeated burst of star formation was inevitable, thus producing the multiple stellar generations we observe today in Galactic globular clusters. All the pieces of the puzzle have been on the table for quite some time. This is our attempt at putting them together.

\section*{Acknowledgments} %We would like to thank the anonymous referee for their constructive comments and suggestions.
APM acknowledges support from the European Research Council (ERC) under the European Union's Horizon 2020 research innovation programme (Grant Agreement ERC-StG 2016, No 716082 'GALFOR', PI: A.P. Milone, http://progetti.dfa.unipd.it/GALFOR).
APM has been supported by MIUR through the FARE project R164RM93XW SEMPLICE (PI: A.P. Milone) and the PRIN program 2017Z2HSMF (PI: L.R. Bedin).

\section*{Data Availability}
No new data were generated or analyzed in support of this research.

\author[0000-0002-7093-7355]{A.\,Renzini}
\author[0000-0002-1276-5487]{A.\,F.\.Marino}
\author[0000-0001-7506-930X]{A.\,P.\,Milone}

\label{lastpage}


\vspace{1 truecm}

\begin{thebibliography}{2014}

\bibitem[\protect\citeauthoryear{ }
{ }]{ }

\bibitem[\protect\citeauthoryear{Adams  et al.}
{2017}]{adams17}Adams, S. M., Kochanek, C. S., Gerke, J. R. et al. 2017, \mnras, 468, 4968

\bibitem[\protect\citeauthoryear{Barbuy, Bica \& Ortolani}
{1998}]{barbuy98}Barbuy, B., Bica, E. \& Ortolani, S. 1998, \aap, 333, 117

\bibitem[\protect\citeauthoryear{Bastian}
{2015}]{bastian15}Bastian, N. 2015, arXiv/1510.01330

\bibitem[\protect\citeauthoryear{Bekki \& Freeman}
{2003}]{bekki03}Bekki, K. \& Freeman, K.C. 2003, \mnras. 346, L11

\bibitem[\protect\citeauthoryear{Bedin et al.}
{2004}]{bedin04}Bedin, L.R., Piotto, G., Anderson, J. et al. 2004, \apj, 605, L125

\bibitem[\protect\citeauthoryear{Bellini et al.}
{2013}]{bellini13}Bellini, A., Piotto, G., Milone, A. P. et al. 2013, \apj, 765, 32

\bibitem[\protect\citeauthoryear{Bellini et al.}
{2017}]{bellini17}Bellini, A., Milone, A. P., Anderson, J. et al. 2017, \apj,  844, 164

\bibitem[\protect\citeauthoryear{Carretta et al.}
{2009}]{carretta09}Carretta, E., Bragaglia, A.,  Gratton, R. G. et al. 2009, \aap, 505, 117

\bibitem[\protect\citeauthoryear{Cox \& Giuli}
{1968}]{cox68}Cox, J.P. \& Giuli, R.T. 1968, Principles of stellar structure (New York: Gordon \& Breach)


\bibitem[\protect\citeauthoryear{D'Antona, Gratton \& Chieffi}
{1983}]{dantona83}D'Antona, F., Gratton, R.G. \& Chieffi, A. 1983, \msait,  54, 173

\bibitem[\protect\citeauthoryear{D'Antona et al.}
{2019}]{dantona19}D'Antona, F.,  Ventura, P., Marino, A.F. et al. 2019, \apj, 871, L19

\bibitem[\protect\citeauthoryear{Decressin et al.}
{2007}]{decressin07}Decressin, T., Meynet, G., Charbonnel, C., Prantzos, N., \& Ekstr\"om, S. 2007, \aap 464, 1029

\bibitem[\protect\citeauthoryear{Dekel et al.}
{2009}]{dekel09}Dekel, A., Birnboim, Y., Engel, G. et al. 2009, \nat, 457, 451

\bibitem[\protect\citeauthoryear{Denissenkov \& Hartwick}
{2014}]{denissenkov14}Denissenkov, P.A. \& Hartwick, F.D.A. 2014, \mnras, 437, L21

\bibitem[\protect\citeauthoryear{de Mink et al.}
{2009}]{demink09}de Mink, S. E., Pols, O. R., Langer, N., \& Izzard, R. G. 2009, \aap, 507, L1

\bibitem[\protect\citeauthoryear{D'Ercole  et al.}
{2010}]{dercole10}D'Ercole, A., D'Antona, F., Ventura, P., Vesperini, E., \& McMillan, S. L. W. 2010, \mnras, 407, 854

\bibitem[\protect\citeauthoryear{Dondoglio  et al.}
{2022}]{dandoglio22}Dondoglio, E., Milone, A. P., Renzini, A. et al. 2022, \apj, 927, 207

\bibitem[\protect\citeauthoryear{Eldridge \& Stanway}
{2022}]{eldridge22}Eldridge, J.J. \& Stanway, E.R. 2022, \araa, ArXiv:2202.01413

\bibitem[\protect\citeauthoryear{Elmegreen}
{2017}]{elmegreen17}Elmegreen, B.G. 2017, \apj, 836, 80

\bibitem[\protect\citeauthoryear{Erb et al.}
{2006}]{erb06}Erb, D. K., Shapley, A. E., Pettini, M., et al. 2006, \apj, 644, 813

\bibitem[\protect\citeauthoryear{Gieles et al.}
{2018}]{gieles18}Gieles, M., Charbonnel, C., Krause, M.G.H. et al. 2018, \mnras, 478, 2461

\bibitem[\protect\citeauthoryear{Gratton, Carretta \& Bragaglia}
{2012}]{gratton12}Gratton, R.G., Carretta, E. \& Bragaglia, A. 2012, \aapr, 20, 50

\bibitem[\protect\citeauthoryear{Harris \& Harris}
{2002}]{harris02}Harris, W.E. \& Harris, G.L.H. 2002, \aj, 123, 3108

\bibitem[\protect\citeauthoryear{Hesser\& Bell}
{1980}]{hesser80}Hesser, J.E. \& Bell, R.A. 1980, \apj. 238, L149

\bibitem[\protect\citeauthoryear{Iben \& Renzini}
{1984}]{iben84}Iben, I.Jr. \& Renzini, A. 1984, \pr, 105, 329

\bibitem[\protect\citeauthoryear{Kashino et al.}
{2017}]{kashino17}Kashino, D., Silverman, J. D., Sanders, D., et al. 2017, \apj, 835, 88,

\bibitem[\protect\citeauthoryear{Kraft}
{1979}]{kraft79}Kraft, R.P. 1979, \araa, 17, 309

\bibitem[\protect\citeauthoryear{Krause et al.}
{2013}]{krause13}Krause, M.,  Charbonnel, C. , Decressin, T.,  Meynet, G. , \& Prantzos, N.  2013, \aap, 552, A121

\bibitem[\protect\citeauthoryear{Lamers et al.}
{2017}]{lamers17}Lamers, H.J.G.L.M., Kruijssen, J.M.D.,  Bastian, N. et al. 2017, \aap, 606, A85
 
\bibitem[\protect\citeauthoryear{Lardo et al.}
{2018}]{lardo18}Lardo, C., Salaris, M., Bastian, N. et al. 2018, \aap, 616, A168

\bibitem[\protect\citeauthoryear{Marino et al.}
{2019a}]{marino19a}Marino, A.F., Milone, A.P., Sills, A. et al. 2019a, \apj, 887, 91

\bibitem[\protect\citeauthoryear{Marino et al.}
{2019b}]{marino19b}Marino, A.F., Milone, A.P., Renzini, A. et al. 2019b, \mnras, 487, 3815

\bibitem[\protect\citeauthoryear{Martell et al.}
{2011}]{martell11}Martell, S. L., Smolinski, J. P., Beers, T. C. \& Grebel, E. K. 2011, \aap, 534, A136

\bibitem[\protect\citeauthoryear{Martocchia et al.}
{2018}]{martocchia18}Martocchia, S., Cabrera-Ziri, I., Lardo, C. et al. 2018, \mnras, 473, 2688

\bibitem[\protect\citeauthoryear{Massari, Koppelman \& Helmi}
{2019}]{massari19}Massari, D., Koppelman, H. H. \& Helmi, A. 2019, \aap, 630, L4

\bibitem[\protect\citeauthoryear{Milone et al.}
{2012}]{milone12} Milone, A. P., Piotto, G., Bedin, L.R. et al. 2012, \apj, 744, 58

\bibitem[\protect\citeauthoryear{Milone et al.}
{2017}]{milone17} Milone, A. P., Piotto, G., Renzini, A. et al. 2017, \mnras, 464, 3636

\bibitem[\protect\citeauthoryear{Milone et al.}
{2018}]{milone18} Milone, A. P., Marino, A.F., Renzini, A. et al. 2018, \mnras, 484, 4046

\bibitem[\protect\citeauthoryear{Milone et al.}
{2019}]{milone19} Milone, A. P., Marino, A.F., Bedin, L.R. . et al. 2019, \mnras, 481, 5098

\bibitem[\protect\citeauthoryear{Milone et al.}
{2020}]{milone20} Milone, A. P., Vesperini, E., Marino, A..F. et al.  2020, \mnras, 492, 5457

\bibitem[\protect\citeauthoryear{Montecinos et al.}
{2021}]{montecinos21}Montecinos, C., Villanova, S., Mu\~nos, C. \& Cortes, C.C. 2021, \mnras, 503, 4336

\bibitem[\protect\citeauthoryear{Ortolani et al.}
{1995}]{ortolani95}Ortolani, S., Renzini, A., Gilmozzi, R., et al. 1995, \nat, 377, 70

\bibitem[\protect\citeauthoryear{Piotto et al.}
{2007}]{piotto07}Piotto, G., Bedin, L. R.,  Anderson, J. et al. 2007, \apj, 661, L53

\bibitem[\protect\citeauthoryear{Piotto et al.}
{2015}]{piotto15}Piotto, G., Milone, A. P., Bedin, L. R., et al. 2015, \aj, 149, 91

\bibitem[\protect\citeauthoryear{Pozzetti, Maraston \& Renzini}
{2019}]{pozzetti19}Pozzetti, L., Maraston, C. \& Renzini, A. 2019, \mnras, 485, 586 

\bibitem[\protect\citeauthoryear{Renzini}
{1983}]{renzini83}Renzini, A. 1983, \msait, 54, 335

\bibitem[\protect\citeauthoryear{Renzini}
{2013}]{renzini13}Renzini, A. 2013, \msait, 84, 162

\bibitem[\protect\citeauthoryear{Renzini et al.}
{2015}]{renzini15}Renzini, A., D'Antona, F., Cassisi, S. et al. 2015, \mnras, 454, 4197

\bibitem[\protect\citeauthoryear{Renzini et al.}
{2018}]{renzini18}Renzini, A., Gennaro, M., Zoccali, M. et al. 2018, \apj, 863, 16

\bibitem[\protect\citeauthoryear{Sana, de Mink \& de Koter}
{2012}]{sana12}Sana, H., de Mink, S. E. \& de Koter, A. 2012, Science, 337, 444

\bibitem[\protect\citeauthoryear{Sander et al.}
{2019}]{sander19}Sander, A.A.C., Hamann, W., Todt, H. et al. 2019, \aap, 621,  A92

\bibitem[\protect\citeauthoryear{Searle \& Zinn}
{1978}]{searle78}Searle, L. \& Zinn, R. 1978, \apj, 225, 357 

\bibitem[\protect\citeauthoryear{Shapiro, Genzel \& F\"orster Schreiber}
{2010}]{shapiro10}Shapiro, K.L.,  Genzel, R. \& F\"orster Schreiber, N.M. 2010, 403, 36 

\bibitem[\protect\citeauthoryear{Schneider et al.}
{2018}]{schneider18}Schneider, F. R. N., Ramirez-Agudelo, O. H., Tramper, F. et al. 2018, \aap, 618, A73

\bibitem[\protect\citeauthoryear{Sollima, Baumgardt \& Hilker}
{2019}]{sollima19}Sollima, A., Baumgardt, H. \& Hilker, M. 2019, \mnras, 485, 1460

\bibitem[\protect\citeauthoryear{Sukhbold et al.}
{2016}]{sukhbold16}Sukhbold, T., Erti, S., Woosley, S.E., Brown, J.M. \& Janke, H.-T. 2016, \apj, 821, 38

\bibitem[\protect\citeauthoryear{Tailo et al.}
{2017}]{tailo17}Tailo, M.,  D'Antona, F., Milone, A. P. et al. \mnras, 465, 1046

\bibitem[\protect\citeauthoryear{Tautvaisiene et al.}
{2021}]{bragaglia21}Tautvaisiene,  G., Drazdauskas, A., Bragaglia, A. et al. 2021, ArXiv2111.10684

\bibitem[\protect\citeauthoryear{Tenorio-Tagle et al.}
{2016}]{tenorio16}Tenorio-Tagle, G.,  Mu\~noz-Tu\~non, C., Cassisi, S. \&  Silich, S. 2016, \apj, 825, 118

\bibitem[\protect\citeauthoryear{Valenti et al.}
{2016}]{valenti16}Valenti, E., Zoccali, M., Gonzalez, O. A. et al. 2016, \mnras, 587, L6

\bibitem[\protect\citeauthoryear{Vanzella et al.}
{2019}]{vanzella19}Vanzella, E., Calura, F.,  Meneghetti, M. et al. 2019, \mnras, 483, 3618 

\bibitem[\protect\citeauthoryear{White \& Rees}
{1978}]{white78}White, S.D.M. \& Rees, M.J. 1978, \mnras, 183, 341

\bibitem[\protect\citeauthoryear{Yusof et al.}
{2022}]{yusof22}
Yusof, N., Hirschi, R.,  Eggenberger, P. et al. 2022, \mnras, 511, 2814




\end{thebibliography}
\end{document}